\documentstyle[11pt,newpasp,twoside]{article}
\input epsf                     
\def\edcomment#1{\iffalse\marginpar{\raggedright\sl#1\/}\else\relax\fi}
\reversemarginpar

\begin{document}
\title{Evidence for gas accretion in galactic disks}
 \author{Thijs van der Hulst}
\affil{Kapteyn Astronomical Institute, P.O. Box 800, NL-9700 AV Groningen, 
the Netherlands}
\author{Renzo Sancisi}
\affil{Osservatorio Astronomico, Via Ranzani 1, I-40127 Bologna, Italy \\
Kapteyn Astronomical Institute, Groningen, the Netherlands}

\begin{abstract}
  
Studies of the H\kern0.1em{\scriptsize I} in galaxies have clearly shown that subtle
details of the H\kern0.1em{\scriptsize I} distribution and kinematics often harbour key
information for understanding the structure and evolution of
galaxies.  
Evidence for the accretion of material has grown over the past many
years and clear signatures can be found in H\kern0.1em{\scriptsize I} observations of
galaxies.  We have obtained new detailed and sensitive H\kern0.1em{\scriptsize I} synthesis
observations of three nearby galaxies which are suspected of
capturing small amounts of H\kern0.1em{\scriptsize I} and show that indeed accretion of
small amounts of gas is taking place in these galaxies.  This could
be the same kind of phenomenon of material infall as observed in the
stellar streams in the halo and outer parts of our galaxy and M~31

\end{abstract}

\section{Introduction}

It has long been an important question how galaxies form and evolve
into the great variety of morphologies we see in the nearby universe.
The common view is that galaxies form hierarchically from small
building blocks through a sequence of merger events. This view has
been motivated by the cold dark matter (CDM) simulations which suggest
that all galaxies form initially as discs (e.g. Baugh, Cole \& Frenk
1996, Klypin et al. 1999, Moore et al. 1999, Steinmetz \& Navarro
2002, Abadi et al. 2003). The signatures of the disks can subsequently
be erased by multiple galaxy mergers (Barnes 1992).  

In this paper we consider the formation of large disks through the
accretion of small companions, the process often indicated as the
nurture of galaxies. Evidence in support of the prediction of this
scenario by cold dark matter simulations has come from several
directions. Zaritsky (1995) and Zaritsky \& Rix (1997) determined 
the star formation rate of several tens of galaxies from stellar population 
studies. They also determined the shapes of these galaxies from photometry 
and found that overall the galaxies with asymmetries in their shapes
have the youngest stellar populations. They acsribed this to recent 
accretion events. Kinematic lopsidedness,
observed in the 2-dimensional H\kern0.1em{\scriptsize I} velocity fields of galaxies
(Verheijen 1997, Swaters et al. 1999) has also been considered as a
result of recent minor mergers. Finally there are at least some twenty
examples of galaxies which in H\kern0.1em{\scriptsize I} show either signs of interactions
and/or have small companions (Sancisi 1999). This suggests that 
galaxies often are in an environment where material for accretion
is available.

\begin{figure}[h]
\plotfiddle{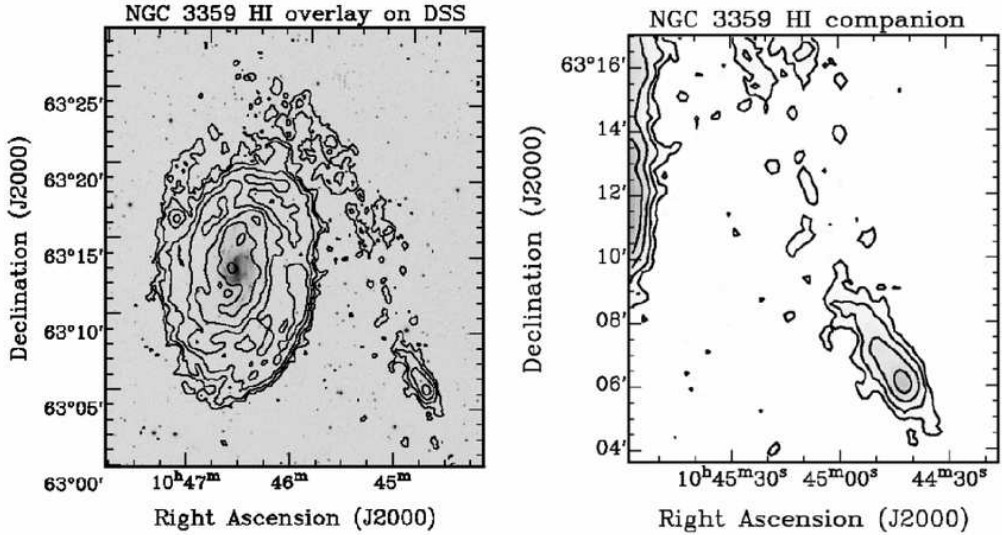}{8.0cm}{0}{92}{92}{-200}{-220}
\caption{H\kern0.1em{\scriptsize I} distribution of NGC~3359 at a resolution of 
30$^{\prime\prime}$ superposed on the digital sky survey 
image (left panel). Contours are 0.1, 0.2, 0.4, 0.8, 1.6, 3.0, and 5.0 
$\times 10^{21}$ cm$^{-2}$. The right panel shows a blow up of the H\kern0.1em{\scriptsize I}
companion.}
\end{figure}
 
\begin{figure}[h] 
\plotfiddle{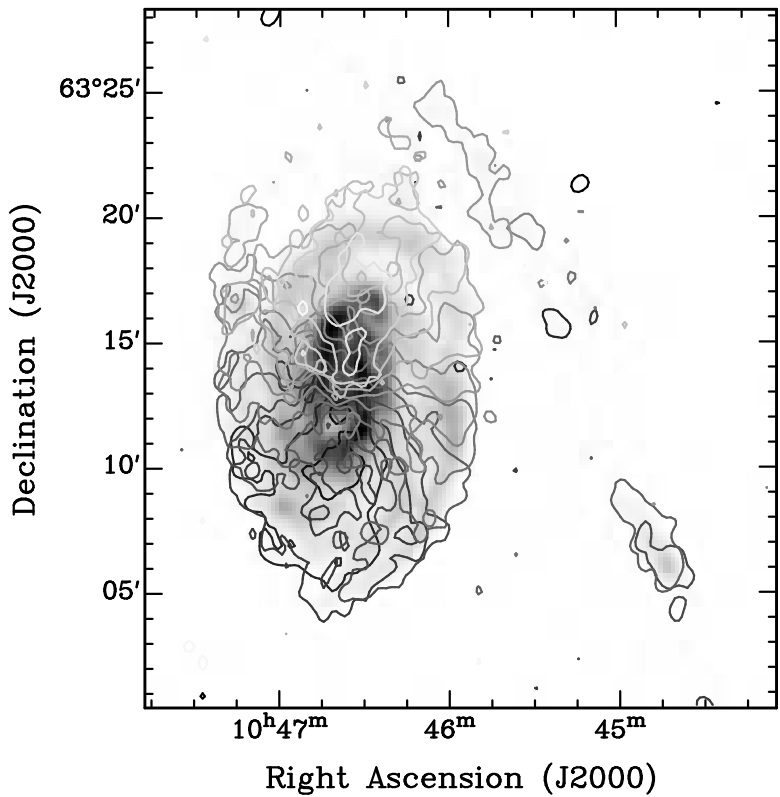}{10.5cm}{0}{58}{58}{-175}{-23}
\caption{Outer contours of the H\kern0.1em{\scriptsize I} emission in individual channels
superposed on the total H\kern0.1em{\scriptsize I} emission in NGC~3359. The low velocities are 
dark grey, the high velocities light grey.}
\end{figure}

Even clearer evidence that accretion events play an important
role has come from studies of the distribution and kinematics of stars
in the Milky Way halo. The discovery of the Sagittarius dwarf
galaxy (Ibata et al. 1994) has been major proof that accretion is
still taking place at the present time. Since such minor merger
remnants retain information about their origin for a long time (Helmi
\& White 2000) studies of the distribution and kinematics of ``stellar
streams'' can in principle be used to trace the merger history of the
Milky Way (Helmi \& de Zeeuw 2001). Such ``stellar streams'' are not
only seen in the Milky Way, but have also been discovered in the Local
Group galaxy M~31 (Ibata et al. 2001, Ferguson et al. 2002,
McConnachie et al. 2003). The substructure in the halo of M~31 is
another piece of clear evidence that minor mergers still take place.

The question now is how to trace such events in more distant galaxies,
where we can not observe individual stars, but require other means of
detecting the signature of accretion.  The use of H\kern0.1em{\scriptsize I} is very powerful
as it can image interactions very effectively by studying the H\kern0.1em{\scriptsize I}
distributions and kinematics of nearby galaxies. Examples can be found
in Sancisi (1999). The improved sensitivity of modern synthesis radio
telescopes brings within reach the detection of faint H\kern0.1em{\scriptsize I} signatures
of accretion events and we expect that new observations of nearby
galaxies will reveal these in the coming decade. To further illustrate
this point we here present a few examples of such signatures:
NGC~3359, NGC~4565 and NGC~6946.

\section{The observations}

All three galaxies have been observed recently with the Westerbork
Synthesis Radio Telescope (WSRT) using the new front-end and
correlator providing a much improved sensitivity.  NGC~3359 and
NGC~4565 were each observed for 12 hours providing a sensitivity of
0.85 mJy/beam for spatial and velocity resolutions of
30$^{\prime\prime}$ and 10 km s$^{-1}$. NGC~6946 was observed for 15
$\times$ 12 hours and reaches a sensitivity of 0.5 mJy/beam for
spatial and velocity resolutions of 60$^{\prime\prime}$ and 5 km
s$^{-1}$.  We will discuss each case individually below.

\subsection{NGC~3359}

NGC~3359 is a nearby barred spiral galaxy (Hubble type SB(rs)c) which
has been observed in H\kern0.1em{\scriptsize I} by Broeils (1992) as part of a study of the
mass distribution of a sample of nearby spiral galaxies. It has a
total mass of $1.2 \times 10^{11}$ M$_{\odot}$ and an H\kern0.1em{\scriptsize I} mass of
$7.5 \times 10^{9}$ M$_{\odot}$ (Broeils \& Rhee, 1997, adjusted for
a Hubble constant of 72 km/s/Mpc).  It has
well developed spiral structure both in the optical and in H\kern0.1em{\scriptsize I}.
Kamphuis \& Sancisi (1994, see also Sancisi 1999) pointed out the
presence of an H\kern0.1em{\scriptsize I} companion which appears distorted and may connect
to the main H\kern0.1em{\scriptsize I} disk of NGC~3359. This observation already suggested
the possibility of witnessing accretion of gas by a large galaxy. Our
new, more sensitive observations are shown in Figure 1 and convincingly
display an H\kern0.1em{\scriptsize I} connection between the distorted H\kern0.1em{\scriptsize I} companion and the
main galaxy. The mass of the H\kern0.1em{\scriptsize I} companion is $1.8 \times 10^{8}$ 
M$_{\odot}$ or 2.4\% of the H\kern0.1em{\scriptsize I} mass of NGC~3359. Also shown in Figure 1
is a blow-up of the H\kern0.1em{\scriptsize I} distribution of the companion which is clearly
distorted and shows a tail pointing towards and connecting with the
outer spiral structure of NGC~3359. No optical counterpart has yet been 
identified.

The velocity structure of the H\kern0.1em{\scriptsize I} companion and the connecting H\kern0.1em{\scriptsize I}
fits in very well with the regular velocity field of NGC~3359. This is
shown in Figure 2 where we display the emission in the individual
channels superposed on the total H\kern0.1em{\scriptsize I} image of NGC~3359.  Contours of
different shades of grey (low velocities are dark, high velocities are
light) denote the outer edge of the H\kern0.1em{\scriptsize I} emission in each of the
velocity channels and thus display the basic kinematics of the H\kern0.1em{\scriptsize I}
without any further analysis of individual velocity profiles. The
regularity of the velocities suggests that the process has been going
on slowly for at least one rotational period which is of the order of
1.7 Gy.

\subsection{NGC~4565}

NGC~4565 is a large edge-on galaxy of Hubble type SAb which was first
observed in H\kern0.1em{\scriptsize I} by Sancisi (1976) in an early search for galaxies with
warped H\kern0.1em{\scriptsize I} disks. Rupen (1991) observed NGC~4565 with much higher
resolution and presented a detailed study of the kinematics and the
warp.  NGC~4565 has a small optical companion 6$^{\prime}$ to the
north of the center of NGC~4565, F378-0021557, which has $7.4 \times
10^{7}$ M$_{\odot}$ of H\kern0.1em{\scriptsize I} compared to an H\kern0.1em{\scriptsize I} mass of $2.0 \times
10^{10}$ M$_{\odot}$ for NGC~4565. Another companion, NGC~4562,
somewhat larger in H\kern0.1em{\scriptsize I} ($2.5 \times 10^{8}$ M$_{\odot}$) and brighter
optically can be found found 15$^{\prime}$ to the south-west of the
center of NGC~4565. The H\kern0.1em{\scriptsize I} distribution, superposed on the DSS is
shown in Figure 3. The asymmetric warp is clearly visible. The warp
sets in at the edge of the optical disk and does exhibit a bit of
apparent thickening of the H\kern0.1em{\scriptsize I} disk visible to the north-west and
south-east of the disk, a result from projection effects along the
line of sight.

\begin{figure}[h]
\plotfiddle{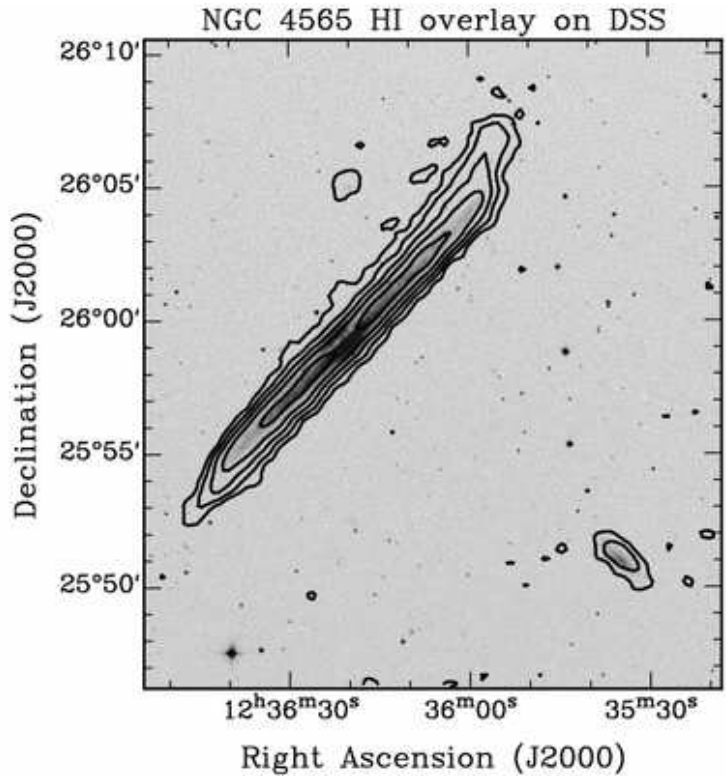}{10.5cm}{0}{90}{90}{-255}{-170}
\caption{H\kern0.1em{\scriptsize I} distribution of NGC~4565 at a resolution of 
30$^{\prime\prime}$ superposed on the digital sky survey 
image (left panel). Contours are 0.4, 0.8, 1.6, 3.2, and 6.4 
$\times 10^{21}$ cm$^{-2}$.}
\end{figure}

\begin{figure}[h] 
\plotfiddle{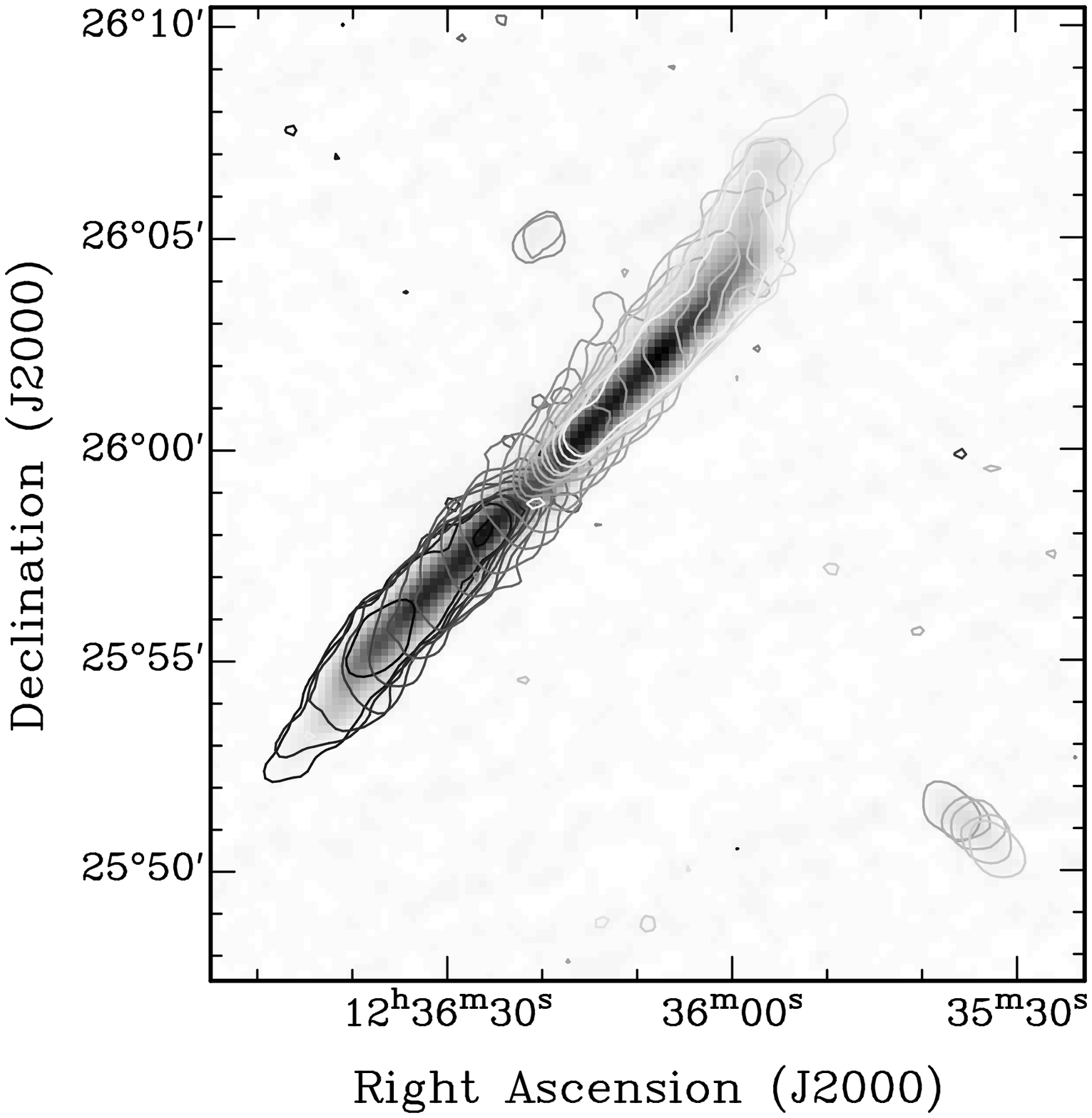}{10.5cm}{0}{58}{58}{-175}{-23} 
\caption{Outer contours of the H\kern0.1em{\scriptsize I} emission in individual channels
superposed on the total H\kern0.1em{\scriptsize I} emission in NGC~4565. The low velocities are 
dark grey, the high velocities light grey.}
\end{figure}

Inspection of individual channel maps brings to light that in addition
to the warp the H\kern0.1em{\scriptsize I} distribution shows additional, low surface
brightness emission to the north of the center, in the direction of
the faint companion F378-0021557. This is best shown in Figure 4 where
we show the outer contours of the H\kern0.1em{\scriptsize I} emission in individual
velocity channels superposed on the total H\kern0.1em{\scriptsize I} distribution of
NGC~4565. In this Figure one clearly sees that there is an additional
extraplanar H\kern0.1em{\scriptsize I} component pointing into the direction of F378-0021557,
suggestive of a connection between the presence of F378-0021557 and
disturbances in the H\kern0.1em{\scriptsize I} disk of NGC~4565. Whether this is recently
accreted material can not easily be verified, but it definitely shows
that there is a component in the H\kern0.1em{\scriptsize I} disk of NGC~4565 which can not be
associated with the warp and is the kind of small (in terms of H\kern0.1em{\scriptsize I}
mass) asymmetry that could be the result of accretion.

\subsection{NGC~6946}

NGC~6946 is a bright, nearby spiral galaxy of Hubble type SAB(rs)cd
which has been studied in H\kern0.1em{\scriptsize I} numerous times (Rogstad et al. 1973,
Tacconi \& Young 1986, Kamphuis 1993). It was in this galaxy that
Kamphuis and Sancisi (1993) found the first evidence for an anomalous
velocity H\kern0.1em{\scriptsize I} component which they associated with outflow of gas from
the disk into the halo as a result of stellar winds and supernova
events. Evidence for such a component is now evident in more galaxies
as discussed by Fraternali et al. (2002, 2003, and also this
volume). A much more detailed study of the anomalous H\kern0.1em{\scriptsize I} and the
structure in the H\kern0.1em{\scriptsize I} disk is being carried out by Boomsma et al. (this
volume, see also Fraternali et al. this volume) on the basis of very
sensitive observations with the WSRT.

\begin{figure}[h]
\plotfiddle{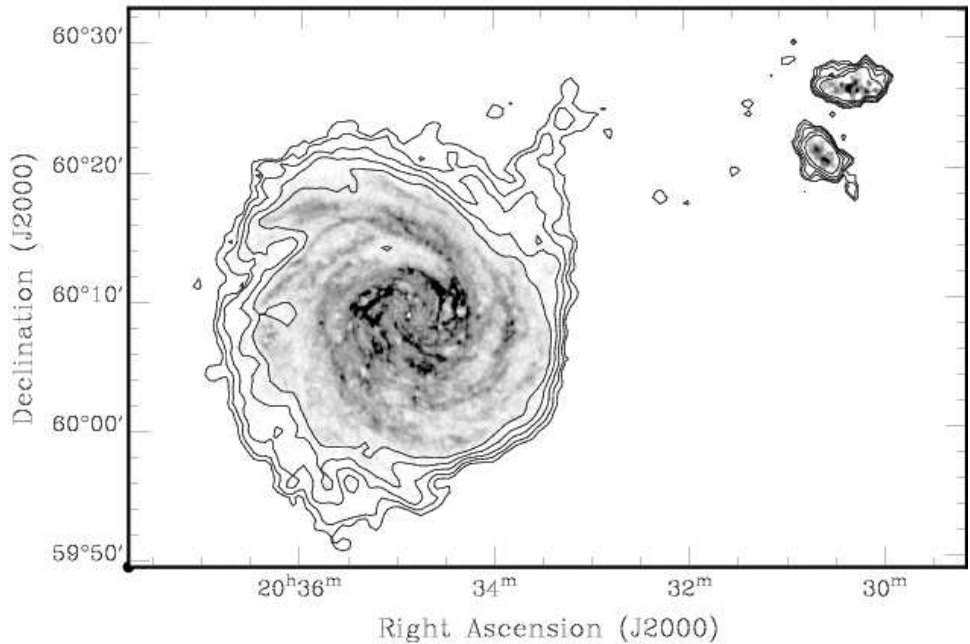}{8.35cm}{0}{73}{73}{-190}{-130}  
\caption{H\kern0.1em{\scriptsize I} distribution in NGC~6946 at a resolution of 
60$^{\prime\prime}$. Contours are 1.3, 2.6, 5.2, 10.4, and 
20.8  $\times 10^{19}$ cm$^{-2}$. }
\end{figure}

Here we concentrate on a low resolution (60$^{\prime\prime}$)
version of these data. Figure 5 shows a total H\kern0.1em{\scriptsize I} image of NGC~6946
down to column density levels of $1.3 \times 10^{-19}$ cm$^{-2}$. To
the west two small companion galaxies can be seen. The most intruiging
feature is the faint whisp to the north-west of the main H\kern0.1em{\scriptsize I} disk of
NGC~6946.  This faint H\kern0.1em{\scriptsize I} extension can only be brought out at this
resolution and appears to form a faint H\kern0.1em{\scriptsize I} filament which blends 
smoothly (also kinematically) with the H\kern0.1em{\scriptsize I} disk of NGC~6946 at a position
some 11$^{\prime}$ (or 19 kpc) south of the tip of the filament. There
is no detectable connection with the two companion galaxies farther to
the west. The spatial and velocity structure of the object are so
regular, yet only connected to the main H\kern0.1em{\scriptsize I} disk at one side that we
prefer an explanation in terms of a tidally stretched, infalling H\kern0.1em{\scriptsize I}
object. So yet another example of accretion of small amount of gas
onto a large H\kern0.1em{\scriptsize I} disk.

Similar examples, though much more massive in H\kern0.1em{\scriptsize I}, are perhaps the
filament discovered in NGC~2403 (Fraternali et al. 2002, 2003 and also
this volume), a long H\kern0.1em{\scriptsize I} filament in M~33 (van der Hulst,
unpublished) and the extraplanar filaments in the northern part of the
H\kern0.1em{\scriptsize I} halo of NGC~891 (Fraternali et al., this volume). 

\section{Concluding remarks}

We have shown three cases with strong evidence for the accretion of
small amounts of H\kern0.1em{\scriptsize I}. These will not be unique. Such faint features
can only be seen in sensitive H\kern0.1em{\scriptsize I} observations as the H\kern0.1em{\scriptsize I} masses
involved are rather modest. We therefore expect that with the
increased sensitivity of modern synthesis radio telescopes, more
examples will be discovered in the coming decade. There probably is a
range of H\kern0.1em{\scriptsize I} masses for these accretion events as is already apparent
from the six cases mentioned here: NGC~891, NGC~2403, NGC~3359,
NGC~4565, NGC~6946 and M~33.

The next question to ask is what the effect of accretion will be on
the disk of the main galaxy. There may very well be a connection with
the star formation activity in galaxies such as NGC~6946 and NGC~2403
and the evidence for gas infall. This then in turn can cause the
observed phenomenon of gas outflows from the active disks as seen in
these galaxies (Boomsma et al. and Fraternali et al. this volume). It is
quite clear that future sensitive and detailed studies of the H\kern0.1em{\scriptsize I} in
nearby galaxies will provide a more complete census of the phenomena
discussed in this paper and enable us to address these issues further
and obtain more definitive answers.




\end{document}